\documentclass[sn-mathphys,Numbered]{sn-jnl}

\usepackage{graphicx}
\usepackage{multirow}
\usepackage{amsmath,amssymb,amsfonts}
\usepackage{amsthm}
\usepackage{mathrsfs}
\usepackage[title]{appendix}
\usepackage{xcolor}
\usepackage{textcomp}
\usepackage{manyfoot}
\usepackage{booktabs}
\usepackage{algorithm}
\usepackage{algorithmicx}
\usepackage{algpseudocode}
\usepackage{listings}

\begin{document}

\title[Mesoamerican duality: $\phi$ and $5/\pi$]{Mesoamerican proportional design and astronomical dualities: rational approximations consistent with $\phi$ and $\pi$ in calendrics and architecture}

\author[1]{\fnm{Gabriel K.} \sur{Kruell}}\email{gabriel.kruell@historicas.unam.mx}
\author[2]{\fnm{Oliver} \sur{L\'opez-Corona}}\email{oliver.lopez@iimas.unam.mx}
\author*[3]{\fnm{Sergio} \sur{Mendoza}}\email{sergio@astro.unam.mx}
\author[4]{\fnm{Pablo} \sur{Padilla}}\email{pablo@iimas.unam.mx}
\author[5]{\fnm{Elvia} \sur{Ram\'{i}rez-Carrillo}}\email{elviarc@ciencias.unam.mx}
\author[3]{\fnm{Sarah\'{\i}} \sur{Silva}}\email{sgarcia@astro.unam.mx}

\affil[1]{\orgname{Universidad Nacional Aut\'onoma de M\'exico},
\orgdiv{Instituto de Investigaciones Hist\'oricas}, 
\orgaddress{\city{Ciudad Universitaria}, \postcode{04510},
\state{Ciudad de M\'exico}, \country{M\'exico}}}

\affil[2]{\orgname{Universidad Nacional Aut\'onoma de M\'exico},
\orgdiv{Investigadores por México (IxM)–Secihti at IIMAS–UNAM},
\orgaddress{\city{Ciudad Universitaria}, \postcode{04510},
\state{Ciudad de M\'exico}, \country{M\'exico}}}

\affil[3]{\orgname{Universidad Nacional Aut\'onoma de M\'exico},
\orgdiv{Instituto de Astronom\'{\i}a},
\orgaddress{\city{Ciudad Universitaria}, \postcode{04510},
\state{Ciudad de M\'exico}, \country{M\'exico}}}

\affil[4]{\orgname{Universidad Nacional Aut\'onoma de M\'exico},
\orgdiv{Instituto de Investigaciones en Matemáticas Aplicadas y Sistemas},
\orgaddress{\city{Ciudad Universitaria}, \postcode{04510},
\state{Ciudad de M\'exico}, \country{M\'exico}}}

\affil[5]{\orgname{Universidad Nacional Aut\'onoma de M\'exico},
\orgdiv{Facultad de Ciencias},
\orgaddress{\city{Ciudad Universitaria}, \postcode{04510},
\state{Ciudad de M\'exico}, \country{M\'exico}}}

\abstract{
Understanding how ancient Mesoamerican societies integrated mathematical ideas into calendrical design and monumental architecture requires approaches that acknowledge their distinct epistemological frameworks. While explicit textual evidence for concepts such as $\pi$ or the golden ratio $\phi$ is absent, numerical patterns embedded in Mesoamerican calendars, iconography, and ritual architecture reveal a coherent system of proportional reasoning grounded in simple integer ratios. Here we show that the numbers 5 and 8, central to Venus, solar calendrical relations and widely represented in Mesoamerican cosmology—generate rational approximations that reproduce, within known construction tolerances, the geometric relations associated with decagonal layouts. Using high-resolution measurements of the Iguana structure at Guachimontones, we demonstrate that its proportions align with integer ratios consistent with those found in the calendrical system and with the practical geometry of the regular decagon, without requiring knowledge of irrational constants. These findings suggest that Mesoamerican builders employed stable proportional modules that harmonized astronomical cycles, cosmological symbolism, and architectural design. This should not be interpreted as a lack of mathematical sophistication, the material record reveals a distinct mathematical tradition in which number, measure, and cosmology were mutually reinforcing elements of cultural knowledge.
}

\maketitle

\section{Introduction}
\label{introduction}

The Mesoamerican calendrical and architectural traditions have long been recognized for their sophistication, yet their mathematical content has often been interpreted through assumptions derived from Western epistemologies. In particular, it has traditionally been claimed, explicitly or implicitly, that concepts such as $\pi$ or the golden ratio $\phi$ were unknown in Mesoamerica because they do not appear in explicit written form. This view stands in contrast with the growing archaeological, archaeoastronomical, and iconographic evidence indicating that Mesoamerican cultures developed a highly structured mathematical sensibility expressed not through abstract treatises, but through the integration of number, proportion, cosmology, and material design.

\begin{figure}
\centering
\includegraphics[width=\textwidth]{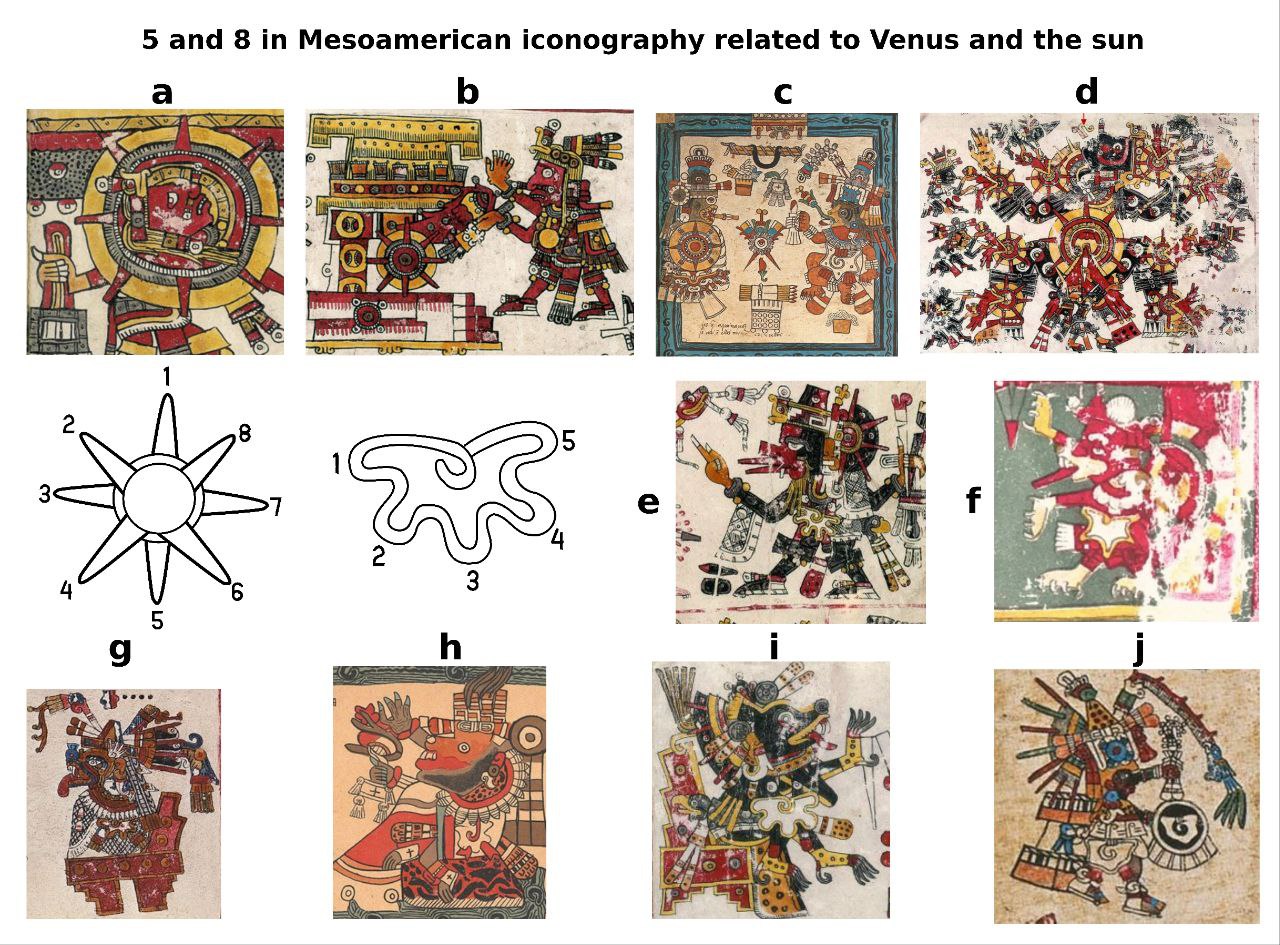}
\caption[Symbolic appearance of numbers $5$ and $8$ in Mesoamerican sources]
{
Solar and Venusian symbolism involving the numbers $5$ and $8$ in Mesoamerican codices.  
These numerological motifs occur in depictions of Tonatiuh, Quetzalcoatl, and Xolotl, and frequently combine five-petaled Venus glyphs with eight-rayed solar disks.  
Their recurrence highlights culturally meaningful dualities that also appear in calendrical design and astronomical relations.
}
\label{figure01}
\end{figure}

A central feature of this intellectual tradition is the recurrent pairing of the numbers $5$ and $8$, deeply embedded in solar and Venusian symbolism (Fig.~\ref{figure01}), and in the astronomical relations governing the Earth–Venus synodic cycle. The equality
\[
365 \times 8 = 584 \times 5,
\]
encodes the fundamental periodicity linking the solar year to the synodic period of Venus. When expressed in dimensionless form, such relations naturally generate simple integer ratios such as $8\!:\!5$, which also appear in Mesoamerican iconography and calendrical numerology.

These integers further resonate with the Fibonacci sequence, whose ratio $8/5 = 1.6$ lies within $1.1\%$ of the golden ratio $\phi = 1.618\ldots$. Archaeological studies of Mesoamerican construction techniques demonstrate consistent proportional accuracy within $1\text{–}3\%$, making such simple ratios archaeologically plausible design choices. It is therefore unnecessary to assume that ancient builders employed irrational constants: rational integer ratios suffice to reproduce, within building tolerances, the geometric relations associated with $\pi$ and $\phi$ in circular or decagonal designs.

The circular ceremonial architecture of Guachimontones, a ceremonial complex of the Teuchitlán tradition located in western Mexico (present-day Teuchitlán, Jalisco), provides a compelling case study. High-resolution orthophotography of the Iguana structure reveals proportions consistent with a decagonal layout governed by rational modules compatible with those appearing in the calendrical system. These observations suggest a mathematical tradition grounded in proportional reasoning, where number, measure, and cosmological order were deeply interconnected. In this framework, architectural geometry, astronomical cycles, and symbolic numerology form a coherent knowledge system without requiring the explicit formulation of modern mathematical constants.


\section{Proportional Design and the Mathematical Structure of Mesoamerican Decagonal Architecture}

The construction of monumental architecture in ancient Mesoamerica relied on cords, poles, and repeated triangulation rather than on written mathematical treatises, yet it consistently achieved a remarkable degree of geometric regularity. Studies of rchitectural orientations across Central Mexico and the Maya region show that many structures were aligned within $0.5^\circ$–$1^\circ$ of intended astronomical or cardinal directions \citep{AveniGibbs1976,AveniHartung1982,Sprajc2015,GonzalezGarciaSprajc2016}. At the spatial scales typical of ceremonial centers, such angular deviations correspond to planimetric errors of only a few percent. These results indicate that sustained precision in linear proportions, often at the $1$–$3\%$ level, was achievable using simple surveying tools when architectural planning was deliberate and symbolically charged.

Within this archaeological context, rational integer ratios such as $3/2$, $4/3$, or $8/5$ become natural candidates for understanding architectural modules. They are easy to construct with cords, stable under iterative layout procedures, and produce geometric forms that, within known tolerances, can be indistinguishable from those generated by irrational constants such as $\phi$ or $\pi$. This is especially relevant for the ratio $8/5 = 1.6$, which is culturally prominent in Mesoamerica through its appearance in the solar--Venus calendrical relation
\[
365 \times 8 = 584 \times 5 = 2920,
\]
and mathematically noteworthy as a close Fibonacci approximation to the golden ratio $\phi \approx 1.618$ (relative error $\approx 1.1\%$). Ratios in this range naturally reproduce the fundamental geometry of the regular decagon, for which the relation between circumradius $R$ and side length $a$ is
\[
\frac{R}{a} = \phi,
\]
and thus require no conceptualization of irrational numbers to yield near-decagonal layouts at archaeological tolerances.

\subsection{The Iguana Structure at Guachimontones}

\begin{figure}
\label{figure02}
\begin{center}
  \includegraphics[width=\textwidth]{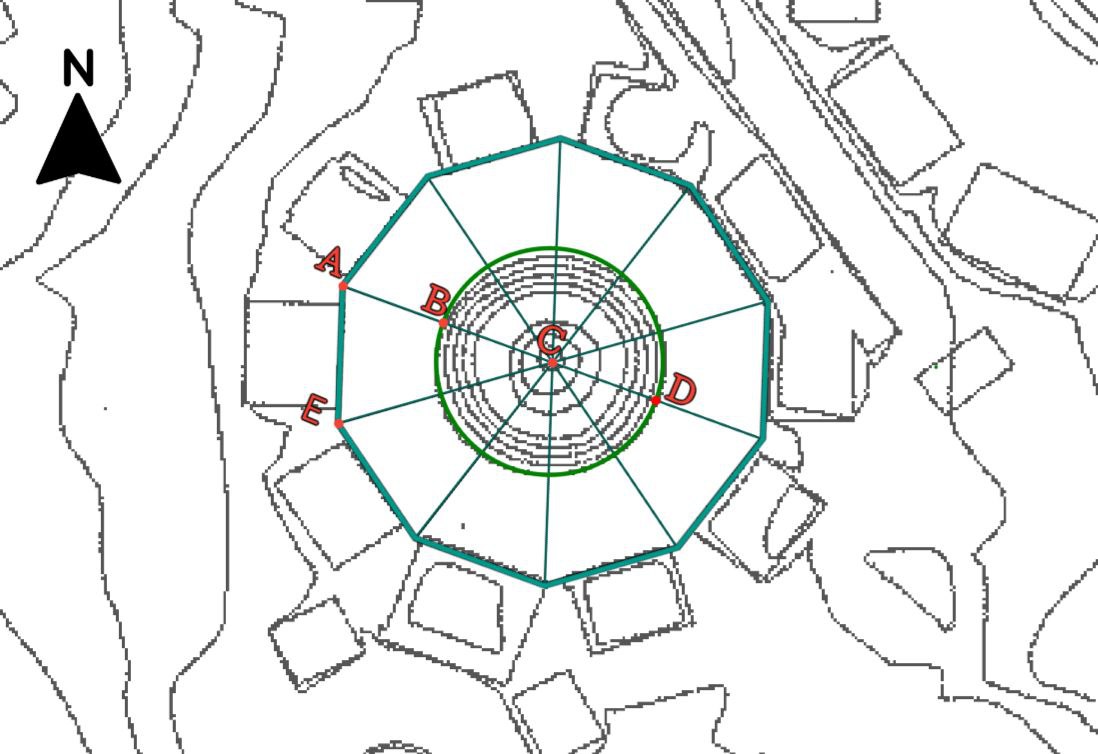}
\caption[Iguana structure at Guachimontones]{The Figure shows a digital
orthophoto of the Iguana
structure at the ``\emph{Recinto Ceremonial de Guachimontones}'',
in Teuchitlan, Jalisco, M\'exico.  The orthophoto was taken
from~\citep{weigand}, with land isocontours at \(1 \text{m}\) intervals.
The figure shows a superimposed best fit regular decagon together with the 
best green circle that bounds the central circular structure.  The
lengths  \( AC \approx BD \approx 38m \) and \( AE \approx 24m \).  In
other words: \( AC / AE \approx BD / AE \approx 38 / 24 = 19/12 =
1.58333 \approx 1.6 \approx 8/5 \approx \phi \), the golden ratio. 
}
\end{center}
\end{figure}

The ceremonial complex of Los Guachimontones, belonging to the Teuchitlán tradition of western Mexico, provides a clear case for evaluating these ideas. The site consists of concentric circular platforms surrounding a central conical pyramid, with documented diameters ranging from roughly $50$ to $120\,\mathrm{m}$ for the primary circles \citep{WeigandBeekman1998, Weigand2004GuachimontonesGuide,Espinoza2021}. The Iguana structure (Fig.~\ref{figure02}) is one of the best preserved of the secondary complexes.

Using a high-resolution orthophoto with $1\,\mathrm{m}$ contour intervals as a geometric reference, we measured the distances
$AC \approx BD \approx 38\,\mathrm{m}$ and $AE \approx 24\,\mathrm{m}$. These yield the empirical ratio
\[
\frac{AC}{AE} = \frac{38}{24} = 1.583\ldots,
\]
a value that lies within $\sim 2\%$ of both the rational approximation $8/5 = 1.6$ and the mathematical golden ratio $\phi \approx 1.618$.
Given the documented construction precision of other Mesoamerican monuments, a deviation of this magnitude is fully consistent with intentional proportioning based on a simple integer module.

\subsection{Uncertainty Quantification via Monte Carlo Simulation}

To assess the sensitivity of this ratio to measurement uncertainty, we performed a Monte Carlo error-propagation analysis. Because the orthophoto provides topographic isocontours at $1\,\mathrm{m}$ intervals, we adopt a conservative uncertainty of $\pm 1\,\mathrm{m}$ for each measured length. In the absence of a full geodetic error model, we treat these uncertainties as independent Gaussian-distributed errors, a standard assumption.

We generated $n = 100{,}000$ realizations by sampling
\[
AC \sim \mathcal{N}(38, 1^2), \qquad
BD \sim \mathcal{N}(38, 1^2), \qquad
AE \sim \mathcal{N}(24, 1^2),
\]
and computing, for each iteration, the average ratio
\[
r_{\mathrm{avg}}
  = \frac{1}{2} \left( \frac{AC}{AE} + \frac{BD}{AE} \right).
\]
The resulting distribution has mean $r_{\mathrm{avg}} = 1.586$ and standard deviation $\sigma \approx 0.072$, with a $95\%$ confidence interval of $[1.452, 1.736]$. The mean differs from $\phi$ by approximately $2\%$, and the interval is fully compatible with the rational ratio $8/5$ under realistic
measurement uncertainty. If the geometric resolution of the orthophoto permits a tighter uncertainty (e.g., $\pm 0.35\,\mathrm{m}$ based on pixel-scale analysis), the confidence interval shrinks proportionally (to approximately $\pm 0.05$), further stabilizing the inferred proportion without requiring
additional field measurements.

\subsection{Implications}

These results support the interpretation of the Iguana complex as a
near-decagonal structure realized through simple proportional modules grounded in Mesoamerican cosmology. Importantly, the match between observed ratios and Fibonacci-derived rational values does not require the use of irrational constants or explicit knowledge of Euclidean decagon formulae. Rather, it illustrates how the culturally salient integers $5$ and $8$, already fundamental
to calendrical reckoning, ritual symbolism, and iconographic dualities, could generate architectural geometries that modern mathematics recognizes as consistent with $\phi$-related proportionality, all within well-documented construction tolerances.

\section{Results}

The proportional analysis of the Iguana structure shows that its measurable geometry aligns with integer ratios found in Mesoamerican calendrical design, especially the pair $(5,8)$ linked to Venus–solar cycles.  
The ratio
\[
AC/AE \approx 1.583
\]
matches the rational value $8/5 = 1.6$ within archaeological precision, and produces a geometric configuration consistent with a near-regular decagon.

Furthermore, approximations such as $5/8$ and $8/5$ reconcile the architectural layout with the calendrical relations governing the 365-day solar year and the 584-day synodic cycle of Venus.  
These findings suggest that architectural and calendrical structures may have been governed by shared proportional modules rather than abstract constants, reinforcing the interpretation of Mesoamerican mathematics as a system of symbolic, cosmological, and practical knowledge integrated through ratio and measure.


\section{Discussion and Final Remarks}
\label{discussion}

This dual calendar system cannot be reduced to a simple system for tracking time, the Mesoamerican calendrical tradition reveals a sophisticated integration of astronomy, proportional reasoning, and cosmological symbolism. The absence of explicit mathematical treatises has often been interpreted as a lack of advanced mathematical concepts. Yet the evidence presented here suggests the contrary: Mesoamerican cultures developed a distinct mathematical tradition grounded in the use of simple, culturally meaningful integer ratios that structured both temporal reckoning and monumental architecture.

Our analysis of the Iguana structure at Guachimontones shows that its proportions are consistent with rational modules found in the Venus–solar calendrical system. These ratios reproduce, within known construction tolerances, geometric forms that modern mathematics associates with the regular decagon and the constants $\pi$ and $\phi$, without implying that such constants were conceptualized abstractly. Instead, these architectural expressions appear to reflect a worldview in which number, cosmology, and material practice were deeply interconnected.

Methodologically, the proportional correspondences identified here are best understood as emergent results of construction practice rather than as products of abstract geometric formalism. Integer-based modules, iterative layout, and radial segmentation constrain architectural plans to a limited set of stable ratios. Within realistic construction tolerances, near-decagonal configurations therefore follow from simple proportional rules, without requiring explicit calculation of geometric constants.

This suggests the need for a broader reconsideration of mathematical thought in ancient Mesoamerica. Rather than approximating Western mathematical frameworks, Mesoamerican knowledge systems appear to follow their own internal logic—one in which measurement, celestial cycles, and symbolic dualities form a coherent and culturally embedded mathematical ontology. Recognizing this distinct tradition enriches our understanding of ancient scientific knowledge and highlights the diverse ways in which human societies have conceptualized and materialized mathematical order.


\backmatter

\bmhead{Acknowledgments}

This work was supported IxM Secihti fellowships program. 

\bmhead{Author Contributions}

All authors contributed to the conceptual development, numerical analysis, and writing of the manuscript.  
All authors discussed the results, revised the text, and approved the final version.

\bmhead{Correspondence}

Correspondence and requests for materials should be addressed to  
Sergio Mendoza (email: \texttt{sergio@astro.unam.mx})  
and  
Oliver L\'opez-Corona (email: \texttt{oliver.lopez@iimas.unam.mx}).

\bmhead{Data Availability}

The datasets supporting the analyses presented in this study are available from the corresponding authors upon reasonable request.

\bmhead{Code Availability}

The codes supporting the analyses presented in this study are available from the corresponding authors upon reasonable request.

\bmhead{Competing Interests}

The authors declare no competing financial or non-financial interests.


\bibliography{mesoamerica-pi-phi}

\end{document}